\documentclass[aps,prb,twocolumn,showpacs,superscriptaddress]{revtex4}
\usepackage{graphicx}
\usepackage{epstopdf}

\begin{document}
\title{Distinguishing $s^{\pm}$ and $s^{++}$ electron pairing symmetries by neutron spin resonance in superconducting NaFe$_{0.935}$Co$_{0.045}$As}

\author{Chenglin Zhang} 
\affiliation{Department of Physics and Astronomy, Rice University, Houston, Texas 77005, USA}
\affiliation{Department of Physics and Astronomy, The University of Tennessee, Knoxville, Tennessee 37996-1200, USA}

\author{H.-F. Li}
\affiliation{J$\ddot{u}$lich Centre for Neutron Science JCNS, Forschungszentrum J$\ddot{u}$lich GmbH, Outstation at Institut Laue-Langevin, Bo$\hat{\imath}$te Postale 156, F-38042 Grenoble Cedex 9, France}
\affiliation{Institut f$\ddot{u}$r Kristallographie der RWTH Aachen, 52056 Aachen, Germany}

\author{Yu Song} 
\affiliation{Department of Physics and Astronomy, Rice University, Houston, Texas 77005, USA}
\affiliation{Department of Physics and Astronomy, The University of Tennessee, Knoxville, Tennessee 37996-1200, USA}

\author{Yixi Su}
 \affiliation{
J\"{u}lich Centre for Neutron Science JCNS-FRM II, Forschungszentrum J\"{u}lich GmbH, Outstation at FRM II, Lichtenbergstrasse 1, D-85747 Garching, Germany 
}

\author{Guotai Tan} 
\affiliation{Department of Physics, Beijing Normal University, Beijing 100875, China}
\affiliation{Department of Physics and Astronomy, The University of Tennessee, Knoxville, Tennessee 37996-1200, USA}

\author{Tucker Netherton} \affiliation{Department of Physics and Astronomy, The University of Tennessee, Knoxville, Tennessee
37996-1200, USA}

\author{Caleb Redding} \affiliation{Department of Physics and Astronomy, The University of Tennessee, Knoxville, Tennessee 37996-1200,
USA}

\author{Scott V. Carr} 
\affiliation{Department of Physics and Astronomy, Rice University, Houston, Texas 77005, USA}
\affiliation{Department of Physics and Astronomy, The University of Tennessee, Knoxville, Tennessee 37996-1200,
USA}

\author{Oleg Sobolev} \affiliation{
Institut f$\ddot{u}$r  Physikalische Chemie, Georg-August-Universit$\ddot{a}$t G$\ddot{o}$ttingen, Tammannstrasse 6, 
37077 G$\ddot{o}$ttingen, Germany}

\author{Astrid Schneidewind} 
\affiliation{Forschungsneutronenquelle Heinz Maier-Leibnitz (FRM-II), TU M$\ddot{u}$nchen, D-85747 Garching,
Germany}

\author{Enrico Faulhaber} \affiliation{Gemeinsame Forschergruppe HZB - TU Dresden, Helmholtz-Zentrum Berlin f\"{u}r Materialien und
Energie, D-14109 Berlin, Germany} \affiliation{Forschungsneutronenquelle Heinz Maier-Leibnitz (FRM-II), TU M$\ddot{u}$nchen, D-85747 Garching,
Germany}

\author{L. W. Harriger} \affiliation{NIST Center for Neutron Research, National Institute of Standards and Technology, Gaithersburg, Maryland 20899, USA}

\author{Shiliang Li} \affiliation{Beijing National Laboratory for Condensed Matter Physics, Institute of Physics, Chinese Academy of
Sciences, Beijing 100190, China}

\author{Xingye Lu} \affiliation{Beijing National Laboratory for Condensed Matter Physics, Institute of Physics, Chinese Academy of
Sciences, Beijing 100190, China}

\author{Daoxin Yao} \affiliation{State Key Laboratory of Optoelectronic Materials and Technology,Sun Yat-Sen University, Guangzhou
510275, China}

\author{Tanmoy Das} \affiliation{Theoretical Division, Los Alamos National Laboratory, Los Alamos, NM, 87545, USA}

\author{A. V. Balatsky} \affiliation{Theoretical Division, Los Alamos National Laboratory, Los Alamos, NM, 87545, USA}

\author{Th. Br$\rm \ddot{u}$ckel}
\affiliation{J$\ddot{u}$lich Centre for Neutron Science JCNS and Peter Gr$\ddot{u}$nberg Institut PGI, JARA-FIT, Forschungszentrum J$\ddot{u}$lich GmbH, 52425 J$\ddot{u}$lich, Germany}

\author{J. W. Lynn} \affiliation{NIST Center for Neutron Research, National Institute of Standards and Technology, Gaithersburg, Maryland 20899, USA}

\author{Pengcheng Dai} \email{pdai@rice.edu} 
\affiliation{Department of Physics and Astronomy, Rice University, Houston, Texas 77005, USA}
\affiliation{Department of Physics and Astronomy, The University of Tennessee,
Knoxville, Tennessee 37996-1200, USA} 
\affiliation{Beijing National Laboratory for Condensed Matter Physics, Institute of Physics,
Chinese Academy of Sciences, Beijing 100190, China}

\begin{abstract} 
A determination of the superconducting (SC) electron pairing symmetry forms the basis for establishing a microscopic mechansim for superconductivity.
For iron pnictide superconductors, the $s^\pm$-pairing symmetry theory predicts the presence of 
a sharp neutron spin resonance at an energy below the sum of hole and electron SC gap energies
($E\leq 2\Delta$) below $T_c$.  On the other hand, the $s^{++}$-pairing symmetry expects a 
broad spin excitation enhancement at an energy above 
$2\Delta$ below $T_c$.
Although the resonance has been observed in iron 
pnictide superconductors at an energy below $2\Delta$ consistent with the $s^\pm$-pairing symmetry, 
the mode has also be interpreted as arising from the $s^{++}$-pairing symmetry
with $E\ge 2\Delta$ due to its broad energy width and the large uncertainty in determining the SC gaps.  
Here we use inelastic neutron scattering to reveal a sharp resonance at $E=7$ meV in 
SC NaFe$_{0.935}$Co$_{0.045}$As ($T_c = 18$ K). On warming towards $T_c$, the mode energy hardly softens while its energy 
width increases rapidly. 
By comparing with 
calculated  spin-excitations spectra within the $s^{\pm}$ and $s^{++}$-pairing symmetries, we conclude that the ground-state resonance in NaFe$_{0.935}$Co$_{0.045}$As is only consistent with
the $s^{\pm}$-pairing, and is inconsistent with the $s^{++}$-pairing symmetry. 
\end{abstract}


\maketitle

\section{introduction}
A determination of the superconducting (SC) electron pairing symmetry is an important step to establish a microscopic theory for high-transition temperature (high-$T_c$) superconductivity \cite{tsuei}.
Since the discovery of iron pnictide superconductors \cite{kamihara,rotter,cwchu}, a peculiar
unconventional pairing state, where superconductivity arises from sign-revised quasiparticle excitations 
between the isotropic hole and electron Fermi pockets near the $\Gamma$ and $M$ points, respectively, has been proposed  \cite{mazin2011n,kuroki,seo}.  A consequence of this so-called  $s^{\pm}$-pairing state is that
the sign-reversed quasiparticle excitations necessitate a sharp resonance in the spin excitations spectra
(termed spin resonance) occurring below
the sum of the hole and electron SC gap energies ($E\leq 2\Delta=\Delta_h+\Delta_e$) at the antiferromagnetic (AF) wave vector ${\bf
Q}$ connecting the two Fermi surfaces [inset in Fig. 1(a)] below $T_c$ \cite{Korshunov,maier}.  
The experimental discovery of
the resonance by neutron scattering in hole and electron-doped BaFe$_2$As$_2$ iron pnictide superconductors 
\cite{christianson,chenglinzhang,castellan,lumsden,schi09,dsinosov09,clester,jtpark,jzhao10,msliu,dsinosov11,dai}
and iron chalcogenide 
Fe(Se,Te) family of materials \cite{hamook,yqiu,lwharriger12}
has provided strong evidence for the $s^{\pm}$-pairing symmetry .
However, the neutron scattering experiments on single crystals of Ba$_{0.67}$K$_{0.33}$Fe$_2$As$_2$ \cite{chenglinzhang},  BaFe$_{2-x}$Co$_{x}$As$_2$ \cite{lumsden,dsinosov09,clester,jtpark},
and Fe(Se,Te) \cite{hamook,yqiu,lwharriger12} superconductors  
have also revealed that the resonance is rather broad in energy [Fig. 1(e)].  In addition, the SC gap energies $2\Delta$ determined from the angle resolved photoemission (ARPES) experiments for hole and electron-doped BaFe$_2$As$_2$ superconductors by different groups \cite{prichard,terashima,evtushinsky,malaeb} can differ dramatically for even the same material, ranging from below to above the neutron spin resonance energy [Fig. 1(e)], and these values 
can also be quite different from those estimated by specific heat \cite{fhardy} and penetration depth measurements \cite{lanluan}.
Because the superconducting gap values for the multi-band 
iron pnictide superconductors are different for different bands \cite{prichard,terashima,evtushinsky,malaeb}, the resonance energy is determined by the 
superconducting gaps in electron and hole bands contributing most to the quasiparticle nesting condition \cite{Korshunov,maier}.   
Therefore, the resonance in some materials may be broad in energy, and this has allowed
 some workers to argue that superconductivity in iron pnictides arises from 
orbital fluctuation mediated $s^{++}$-pairing superconductivity \cite{evtushinsky,onari10,onari11,nagai,borisenko}, where one expects a broad
spin excitation enhancement (neutron spin resonance)
at an energy of $E\ge 2\Delta$ below $T_c$ \cite{onari10,onari11,nagai}. 
 Given the current debates concerning the universality of the electron pairing symmetry in iron-based superconductors \cite{borisenko,hirschfeld,chubukov,jph13,RMF13,dagotto},
it is important to compare spin excitations in different classes of iron superconductors and determine if the resonance in these 
materials agrees with predictions of the $s^{\pm}$- and $s^{++}$-pairing symmetries.

In this article, we present inelastic neutron scattering results on single crystals of SC
NaFe$_{0.935}$Co$_{0.045}$As with $T_c=18$ K [Figs. 1(a)-1(d)].  In the normal state, the imaginary part of the
dynamic susceptibility, $\chi^{\prime\prime}({\bf Q},E)$, at the AF wave vector increases linearly with increasing energy $E$.  
Upon entering into the 
low-temperature SC state, a spin gap opens below 5.5 meV and a sharp
neutron spin resonance appears at $E=7$ meV with an energy width of $1.5\pm 0.1$ meV [Figs. 2 (d) and 2(f)].
On warming to $T_c$, the resonance broadens in energy width but its central position 
does not follow the decreasing SC gap energy, different from the earlier work on electron-doped BaFe$_2$As$_2$ \cite{dsinosov09}.
By comparing the neutron scattering results with  a random-phase
approximation (RPA) spin-susceptibility calculation within the $s^{+-}$ and $s^{++}$ pairings, we find
that our data are consistent with the  $s^{+-}$ symmetry in the low-temperature SC state and cannot be explained by the $s^{++}$ pairing symmetry.

\begin{figure}[t] \includegraphics[scale=.4]{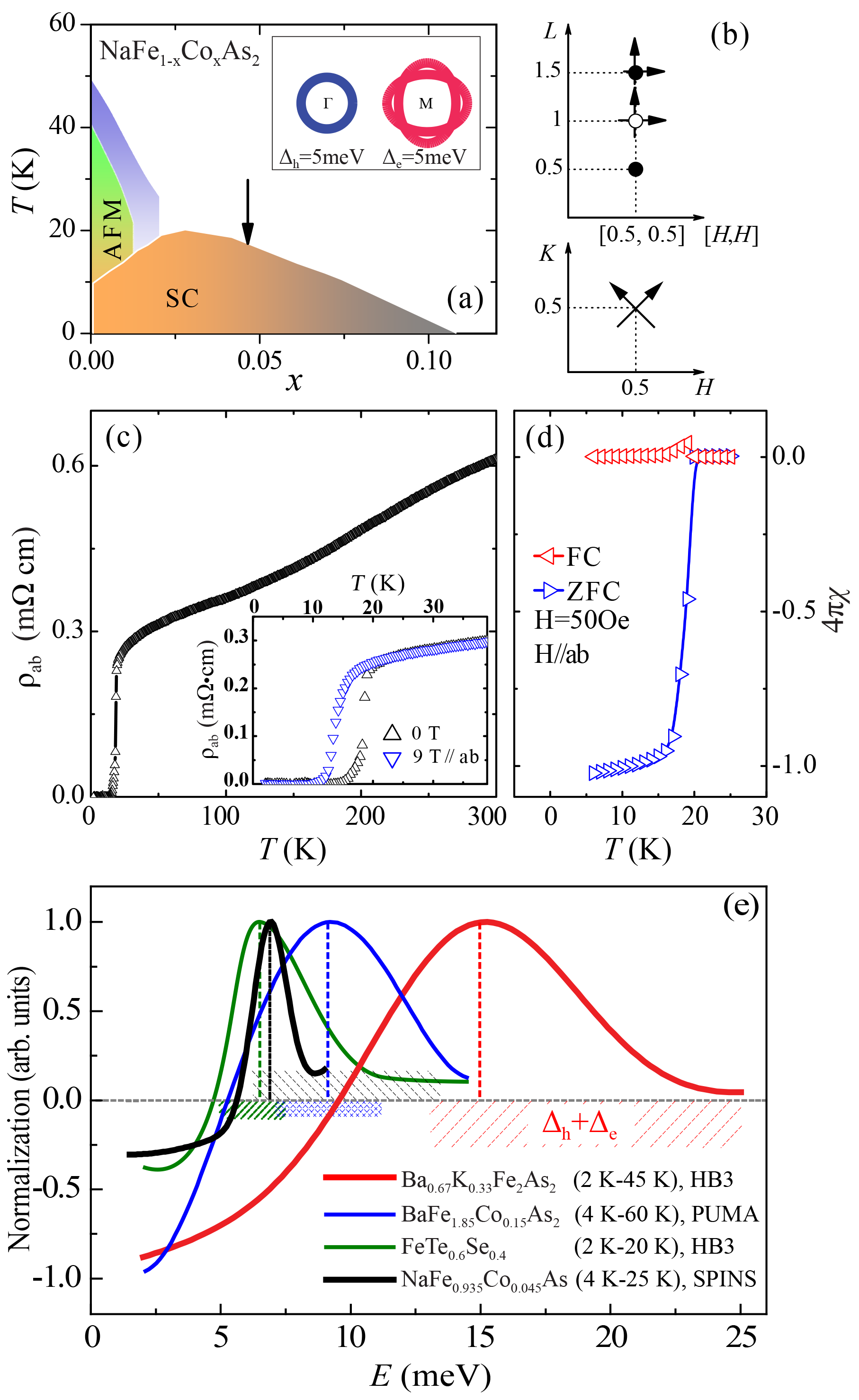} 
\caption{
(Color online) 
{Phase diagram, Fermi surfaces, 
reciprocal space, transport, and susceptibility measurements.} 
(a) The electronic phase diagram of NaFe$_{1-x}$Co$_x$As, where the arrow indicates the Co-doping level 
of our samples.  Inset shows the hole and electron Fermi pockets near $\Gamma$ and $M$ positions, respectively (Ref. \cite{mazin2011n}).
 (b) Reciprocal space probed in the present experiment. The scan directions are shown as arrows.
 (c) Temperature dependence of the in-plane resistivity $\rho_{ab}$ in NaFe$_{0.935}$Co$_{0.045}$As. The inset displays the
 low-T resistivity measured in zero-field and 9 T. 
 (d) The temperature dependence of the bulk susceptibility measured by DC magnetic susceptibility. (e) 
 The schematics of neutron spin resonance for optimally hole (Ref. \cite{chenglinzhang}, red solid line) 
 and electron (Ref. \cite{dsinosov09}, blue solid line) iron pnictide superconductors.  The green solid line shows 
 results for FeTe$_{0.6}$Se$_{0.4}$ superconductor \cite{lwharriger12} 
 and black solid lines show results from the 
 present work on NaFe$_{0.935}$Co$_{0.045}$As.  The red, blue, and black dashed regions show the range of $2\Delta$ as determined from 
 different ARPES and other experiments.  In all cases, the resonance can be either below or above the $2\Delta$.
 }
\end{figure}

\section{Results} 

We carried out inelastic neutron scattering experiments on the thermal (PUMA) and cold (PANDA) 
triple-axis spectrometers at the FRM-II, TU M\"{u}chen, Germany \cite{schi09}, and also on the SPINS cold triple-axis spectrometer 
at NIST Center for Neutron Research, Gaithersburg, Maryland \cite{chenglinzhang}.
For the experiments, we coaligned 5 pieces of self-flux grown NaFe$_{1-x}$Co$_x$As single crystals with a 
total mass of 5.5 g (mosaic about $3^\circ$) for PUMA and PANDA experiment.  Samples are loaded inside a 4 K displex controlled by a calibrated
temperature sensor.

For SPINS measurements, we used $\sim$10 g of single crystals with mosaic less than 3$^\circ$.
The chemical compositions of the samples are determined as
Na$_{1.06}$Fe$_{0.935}$Co$_{0.045}$As by inductively coupled plasma atomic-emission spectroscopy, which 
has an accuracy of about 2\%.  Samples from different batches show almost identical chemical composition, which 
we denote as NaFe$_{0.935}$Co$_{0.045}$As.  
The wave vector ${\bf Q}$ at ($q_x$,$q_y$,$q_z$) in \AA$^{-1}$ is
defined as (\textit{H},\textit{K},\textit{L}) = ($q_xa/2\pi$,$q_ya/2\pi$,$q_zc/2\pi$) reciprocal lattice unit (r.l.u) using the
tetragonal unit cell (space group P4/nmm, $a = 3.921$ {\AA}, $c = 6.911$ {\AA} at 5 K).   We used focusing pyrolytic graphite (PG) monochromator and analyzer with fixed final energies of $E_f=14.7$ meV
and $E_f=5$ meV at PUMA and PANDA, respectively.  For SPINS measurements, we used $E_f=5$ meV with flat monochromator and analyzer.
Both the $[H,H,L]$ and $[H,K,0]$ scattering zones have been used in the experiments and
the scan directions are marked in Fig. 1(b). To characterize the samples, we have 
carried out resistivity and DC magnetic susceptibility measurements using commercial physical property measurement system and SQUID magnetometer.
Based on the early neutron diffraction measurements \cite{slli09}, AF 
Bragg peaks and low-energy spin excitations are expected to occur around the $(0.5,0.5,L)$ positions with $L=0.5,1.5,\cdots$ [Fig. 1(b)].

Figure 1(c) plots the in-plane resistivity $\rho_{ab}$ measurement at zero field which gives $T_c = 18$ K.
The inset shows the magnetic field dependence of $\rho_{ab}$ at 0 and 9-T, indicating a field-induced $T_c$
suppression of $\sim$2 K.  Figure 1(d) shows the magnetic susceptibility measurements on the sample again showing a $T_c=18$ K. 
Given the known electronic phase diagrams of NaFe$_{1-x}$Co$_x$As \cite{Parker,xianhui,guotai}, it is clear that 
our NaFe$_{0.935}$Co$_{0.045}$As samples are in the slightly 
overdoped regime and do not have static AF order coexisting with superconductivity [Figs. 1(a)].
Our elastic neutron diffraction scans through the AF Bragg peak positions are featureless and thus 
confirm this conclusion (see Fig. 7 in the appendix for details).
Figure 1(e) shows schematics of the resonance for optimally hole \cite{chenglinzhang} and electron \cite{lumsden,dsinosov09,clester,jtpark} doped iron pnictides,
as well as iron chalcogenide 
Fe(Se,Te) \cite{hamook,yqiu,lwharriger12},
and the ranges of SC gaps for these materials as determined from ARPES and other techniques \cite{prichard,terashima,evtushinsky,malaeb,fhardy,lanluan,borisenko}.
For comparison, we also show the SC gaps determined from ARPES for NaFe$_{0.95}$Co$_{0.05}$As \cite{Liu_arpes,thirupathaiah}.  As we can see, the ARPES measurements from two groups 
on NaFe$_{0.95}$Co$_{0.05}$As have yielded SC gaps different by a factor of two.

\begin{figure}[t] \includegraphics[scale=.35]{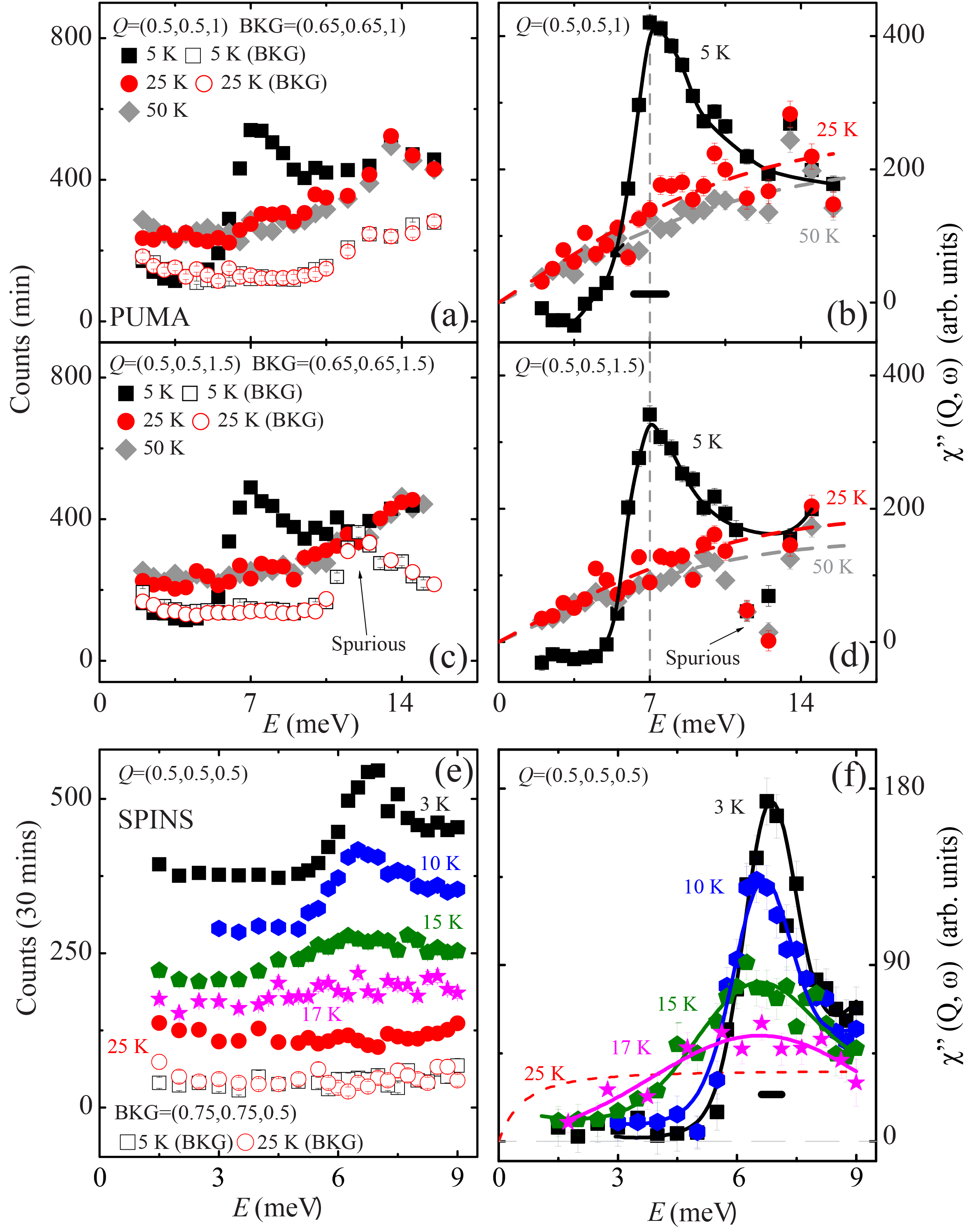} 
\caption{
(Color online) 
{Energy scans at the AF wave vector.} (a) and (c) Energy scans  
at ${\bf Q}=(0.5,0.5,1)$ and ${\bf Q}=(0.5,0.5,1.5)$, respectively, at 5, 25, and 50 K on PUMA. 
The
background was taken at ${\bf Q}=(0.65,0.65,1)$ and ${\bf Q}=(0.65,0.65,1.5)$, respectively, at 5 and 25 K. (b) and (d) are the
corresponding $\chi^{\prime\prime}(Q,E)$.(e) Energy scans at ${\bf Q}=(0.5,0.5,0.5)$ 
at 3, 10, 15, 17, and 25 K on SPINS.  The background scans at ${\bf Q}=(0.75,0.75,0.5)$ 
between 5 and 25 K are also shown.  For clarity, the energy scans at ${\bf Q}=(0.5,0.5,0.5)$ were shifted above 
background data by 30 at 25 K, 100 at 17 K, 150 at 15 K, 230 at 10 K, and 330 at 3 K.
(f) is the corresponding $\chi^{\prime\prime}(Q,E)$. The solid and dashed lines are guides to
the eyes. The horizontal bars indicate instrumental energy resolution. } 
\end{figure}

In previous neutron scattering 
work on electron doped BaFe$_{2-x}$Ni$_x$As$_2$ pnictide superconductors \cite{schi09,jtpark}, the neutron spin resonance was found to be dispersive, occurring at slightly different energies for different $c$-axis wave vector transfers.  To see if this is also the case for spin excitations in NaFe$_{0.935}$Co$_{0.045}$As,
we carried out constant-${\bf Q}$ scans at wave vectors ${\bf Q}=(0.5,0.5,1)$ and $(0.5,0.5,1.5)$
below and above $T_c$ on PUMA. 
While the background scattering (BKG) taken at ${\bf Q} =
(0.65, 0.65, 1)$ and $(0.65, 0.65, 1.5)$ showed no change below and above $T_c$ [Figs. 2(a) and 2(c)],
the scattering at the in-plane AF wave vector revealed dramatic changes across $T_c$.
In the normal state ($T=25$ K), the scattering above BKG is featureless and increases with increasing energy. Upon entering into the SC state ($T=5$ K), a spin gap forms below $\sim$5.5 meV and a sharp resonance appears at $E=7$ meV [Figs. 2(a) and 2(c)]. 
The corresponding $\chi^{\prime\prime}(Q,E)$, obtained by subtracting the BKG and correcting for the Bose population factors using $\chi^{\prime\prime}(Q,E)=[1-\exp(-E/k_BT)]S(Q,E)$,
are shown in Figs. 2(b) and 2(d).  Inspection of Figs. 2(a)-2(d) reveals that the resonance exhibits no $c$-axis dispersion and has an energy width of $\sim$3 meV. 
Figures 2(e) and 2(f) show similar scans on SPINS, which reveal a 5 meV spin gap and a sharp resonance at $E=6.8\pm0.1$ meV 
with an energy width of $\sim$1.5 meV in the SC state at 3 K.  
This is much narrower than the energy widths of the resonances in the hole and electron-doped BaFe$_2$As$_2$ [Fig. 1(e)] \cite{christianson,chenglinzhang,castellan,lumsden,schi09,dsinosov09,clester,jtpark,jzhao10,msliu}.
On warming to 10, 15, and 17 K, the resonance energy widths become broader, but its peak position remains almost unchanged up to $T=0.94T_c=17$ K [Fig. 2(f)].
This is different from earlier work on the temperature dependence of the resonance energy in BaFe$_{1.85}$Co$_{0.15}$As$_2$ ($T_c=25$ K) \cite{dsinosov09}, but somewhat similar
to the temperature dependence of the resonance in Fe(Se,Te) family of materials \cite{yqiu,lwharriger12}.

\begin{figure}[t] \includegraphics[scale=.35]{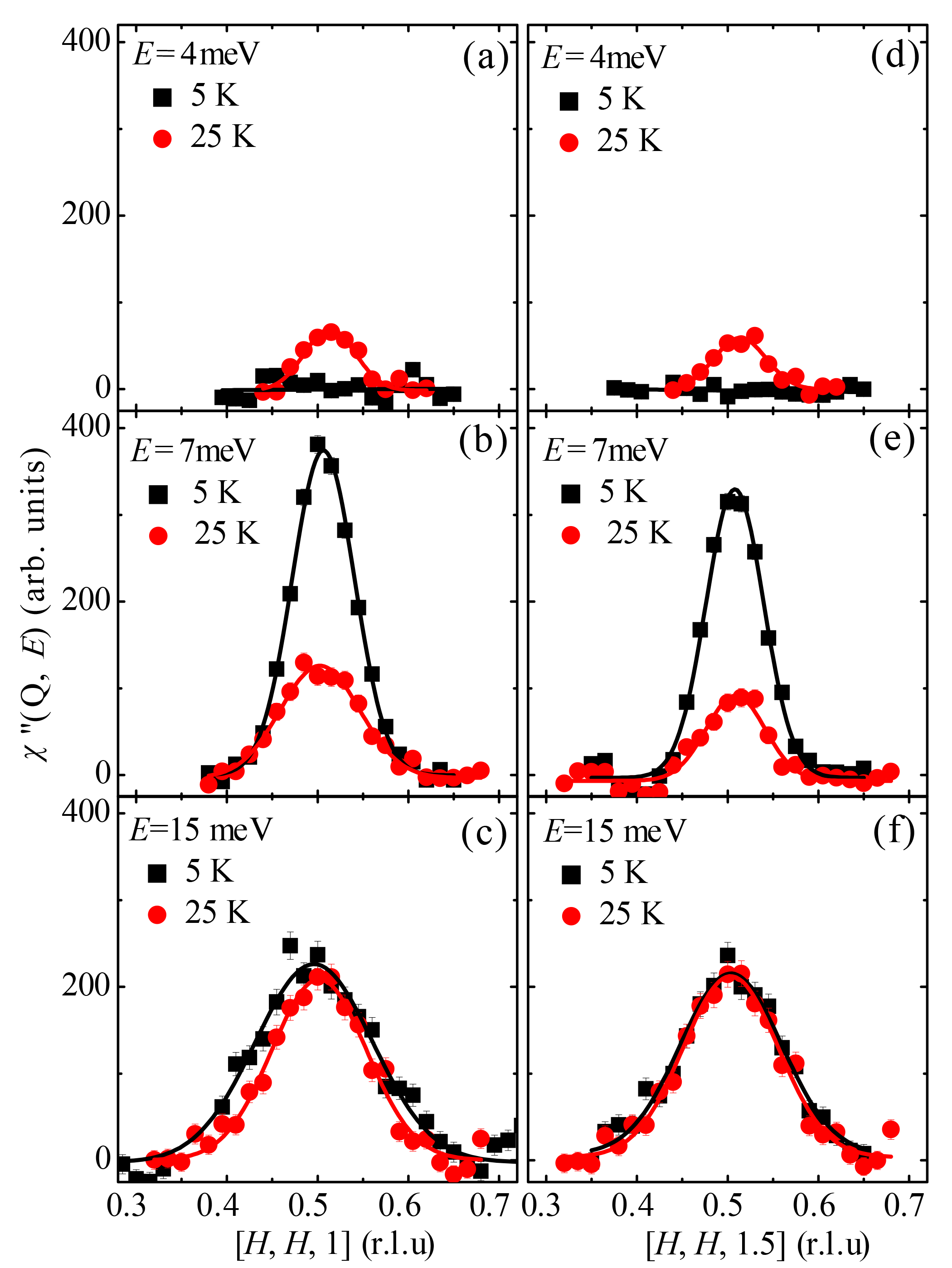} 
\caption{
(Color online)
{Constant-energy scans at different $L$ values across $T_c$.} $\bf Q$ scans along the $[\textit{H},\textit{H},L=1,1.5]$ directions
 below and above $T_c=18$ K at different energies: (a,d) in the gap (4 meV), (b,e) at the resonance energy (7 meV), and (c,f) well 
 above 2$\Delta$ (15 meV). The solid lines are fits to Gaussians. Data are from PUMA.}
\end{figure}

To confirm the SC spin gap and determine the wave vector dependence of the resonance, we carried out
constant-energy scans at $E=4, 7$, and 15 meV below and above $T_c$. 
Figures 3(a-c) and 3(d-f) show $\chi^{\prime\prime}(Q,E)$ along the $[H,H,1]$ and
$[H,H,1.5]$ directions, respectively. 
In the SC state at $T=5$ K, $\chi^{\prime\prime}(Q,E)$ is featureless at $E=4$ meV and thus confirms
the presence of a spin gap. For other excitation energies, the
scattering profiles can be fitted by Gaussians on linear BKG. Taking the Fourier transforms of the fitted Guassian peaks along the $[H,H,1]$ direction, we find that the in-plane spin-spin correlation lengths at the energy of the resonance are $\xi=24\pm 1$ \AA\ at
25 K and $\xi=30\pm 1$ \AA\ at 5 K.  Along the $[H,H,1.5]$ direction, $\xi=32\pm 3$ \AA\ are unchanged from
25 K to 5 K. Increasing the excitation energy to $E=15$ meV ($>2\Delta$), there are no observable differences 
in scattering intensity and spin correlation lengths ($\xi=20\pm 1$ \AA) below and above $T_c$

\begin{figure}[t] \includegraphics[scale=0.38]{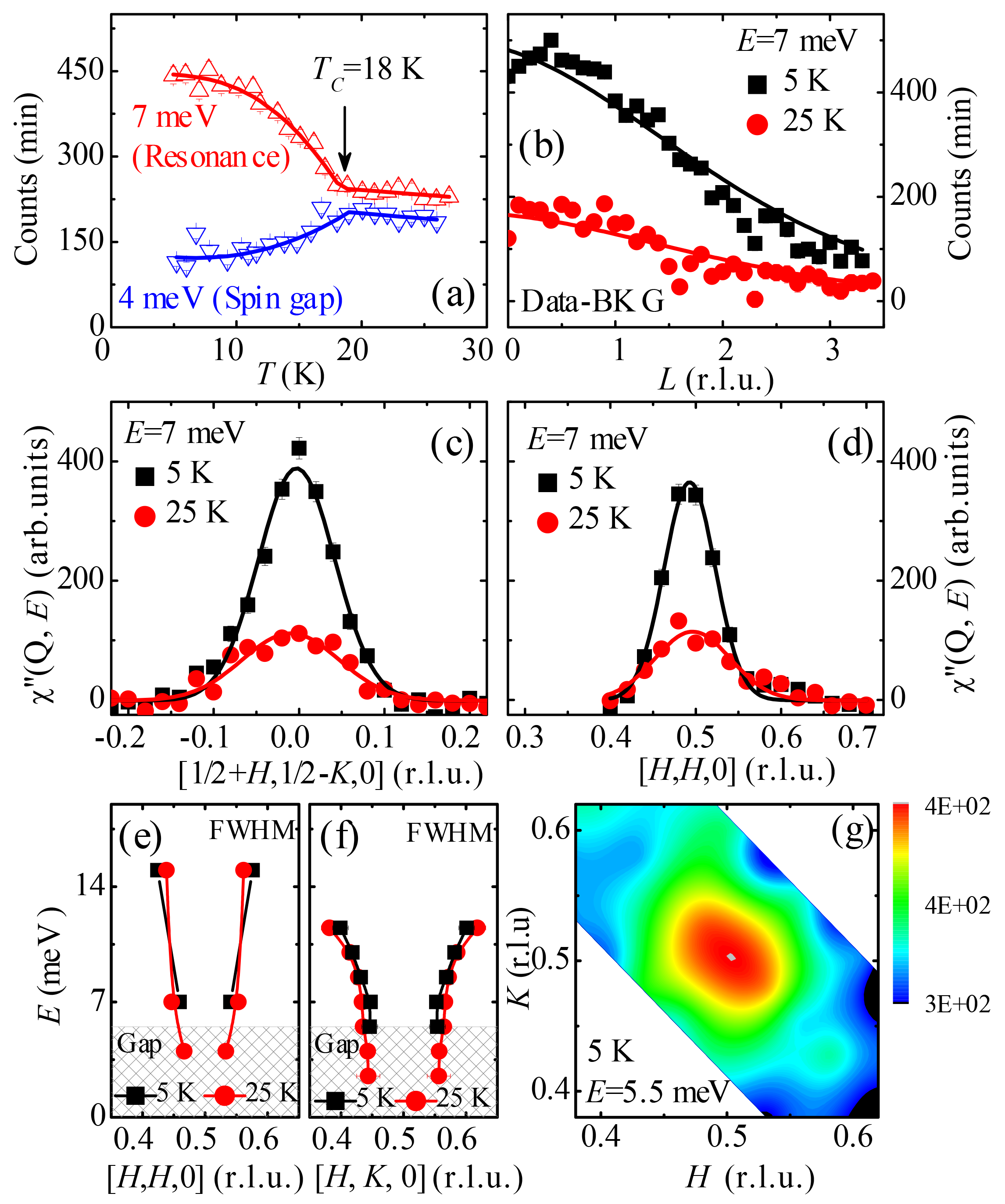} 
\caption{
(Color online)
{Temperature dependence of the resonance and spin gap.} (a) Temperature dependence of the scattering at ${\bf
Q}=(0.5,0.5,1.5)$ at $E=4$, and 7 meV.
The solid lines show the order parameter fits
by using $\textit{I}=\textit{I}_o+\textit{K}(1-(\textit{T}/\textit{T}_c))^\beta$ yielding $T_c=18.8$ K for both. (b)
Background subtracted constant energy scans with \textit{E} = 7 meV along the $[0.5, 0.5, \textit{L}]$ direction at 5 and 25 K. Background was
taken at ${\bf Q}=(0.65,0.65,\textit{L})$. The solid lines are the square of the Fe$^{2+}$ form factor. (c) and (d) ${\bf Q}$-scans 
below and above $T_c$ along the $[1/2+\textit{H}, 1/2-\textit{H},0]$ and
$(\textit{H},\textit{H},0)$ directions, respectively. (e) and (f) The dispersions of spin excitations below and above $T_c$ 
along the $[H,H,0]$ and $[H-\delta,K+\delta,0]$ directions, respectively.
 The shaded area indicates the size of the spin gap in the SC state.
 (g) The in-plane wave vector profile of the $E=5.5$ meV spin excitations in the SC state. Data are from PUMA.} \end{figure}

Figure 4(a) shows the temperature dependence of the scattering at the AF wave vector 
${\bf Q} = (0.5, 0.5, 1.5)$ for the resonance ($E=7$ meV) and spin gap ($E=4$ meV) energies.
While the intensity increases dramatically below $T_c$ at the resonance energy, it decreases at $E=4$ meV
signaling the opening of a SC spin gap.  To test 
if spin excitations in NaFe$_{0.935}$Co$_{0.045}$As are indeed two-dimensional 
in reciprocal space like in the case of Co-doped BaFe$_2$As$_2$ \cite{lumsden}, we show in Fig. 4(b) BKG subtracted
constant-energy scans along the $[0.5,0.5,L]$ direction
at the resonance energy ($E=7$ meV) below and above $T_c$.
The monotonic decrease of the scattering with increasing $L$ is consistent with the square of the
Fe$^{2+}$ magnetic form factor, thus confirming the two-dimensional and magnetic nature of the resonance.

\begin{figure}[t] \includegraphics[scale=0.33]{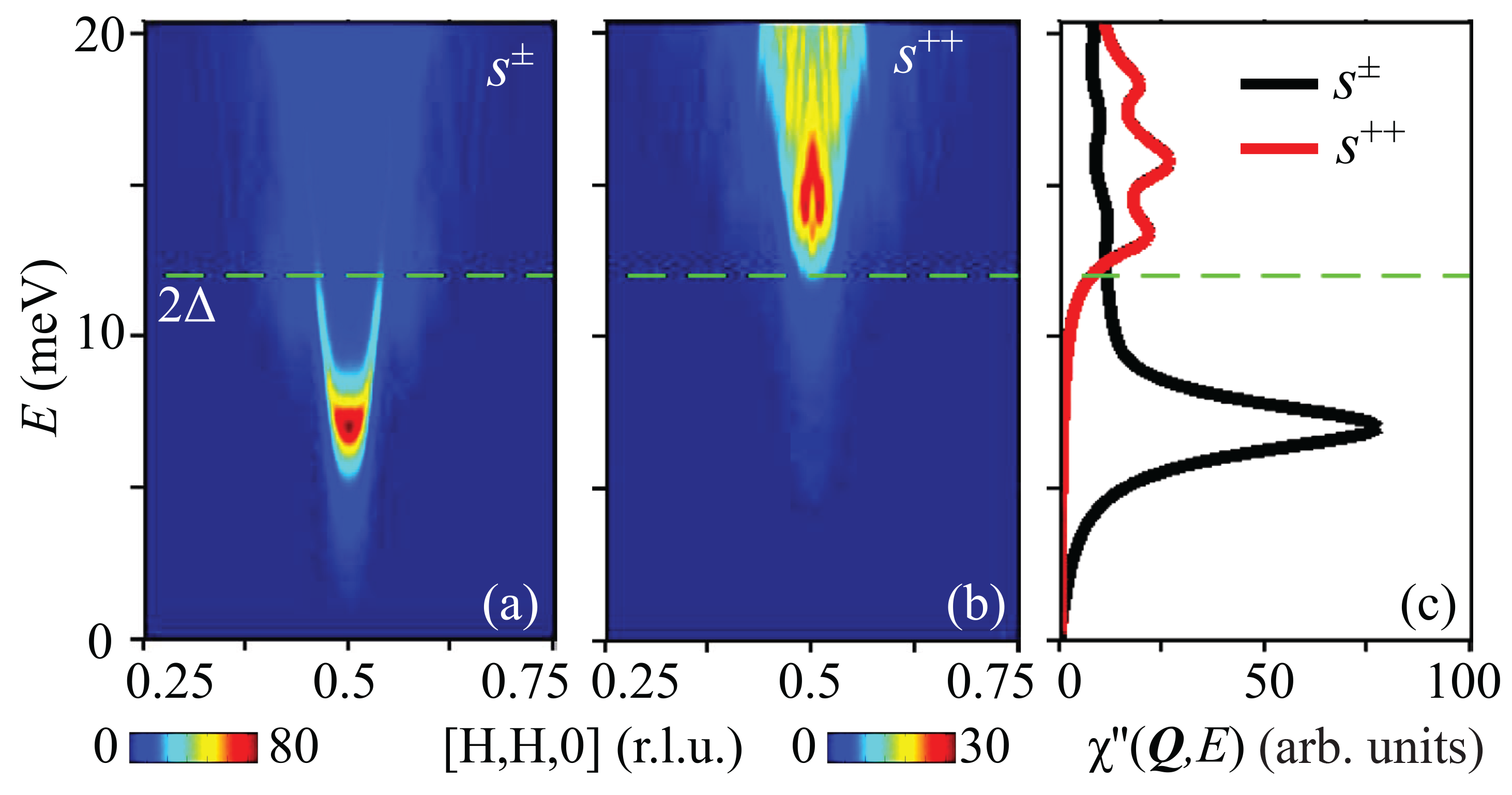} 
\caption{ 
(Color online) {RPA calculated resonance for $s^{\pm}$ and $s^{++}$ pairing symmetries.} (a) Computed spin-excitation
spectrum for the $s^{\pm}$ pairing channel. (b) Same but for $s^{++}$ pairing symmetry. (c) $\chi^{\prime\prime}(\omega)$ at
the AF wave vector ${\bf Q}$ for both these cases. The horizontal lines mark the $2\Delta_0$ line, below and above
which the resonance occurs in the two cases, respectively. Both calculations are performed with fixed intrinsic broadening of 1~meV, SC
gap of $\Delta_0$=6~meV, and Coulomb interaction $U$=1.6~eV.
} \end{figure}

In the optimally hole \cite{chenglinzhang} and electron-doped BaFe$_2$As$_2$ \cite{clester,jtpark,msliu},
the resonances form longitudinally and transversely elongated ellipses, respectively. Since 
NaFe$_{0.935}$Co$_{0.045}$As belongs to the electron-doped iron pnictide superconductor, we mapped out 
spin excitations in the $[H,K,0]$ scattering plane on PUMA. Figures 4(c) and 4(d) show
constant-energy scans at $E=7$ meV along the transverse $[1/2+H,1/2-H,0]$ and longitudinal $[H,H,0]$ directions [perpendicular and parallel to
the in-plane AF ordering wave vector ${\bf Q}=(0.5,0.5,0)$], respectively, below and above $T_c$.  Although there is no evidence for transverse incommensurate magnetic scattering as in the case of LiFeAs \cite{qureshi}, the spin resonance in NaFe$_{0.935}$Co$_{0.045}$As is considerably broader along the transverse direction than that of the longitudinal direction. Figure 4(g) shows the two-dimensional image of the resonance in the SC state, further confirming the results of Fig. 4(c) and 4(d).

Figure 4(e) and 4(f) summarizes the full-width-half-maximum (FHWM) 
of the low-energy spin excitations along the longitudinal and transverse directions below and above $T_c$, respectively.  
In the normal state, spin excitations in NaFe$_{0.935}$Co$_{0.045}$As are gapless, comparing with 
the $\sim$10 meV anisotropy gap in spin waves of the undoped NaFeAs \cite{Inosov}.
The data points are the FWHM of spin excitations along the two high symmetry directions.
On cooling to below $T_c$, the effect of superconductivity  is to open a low-energy spin gap and 
concurrently form a neutron spin resonance.  The dispersions of the spin excitations are essentially unaffected by superconductivity.

\section{Discussion and Conclusions}
To compare with the experiment, we have performed RPA spin-susceptibility calculation in the SC state, using the five-orbital
tight-binding model taken from Ref.~\cite{Graser_TB}. The details of the calculations can be found in
Refs.~\cite{Dasresonance} and appendix.  Results for the $s^{+-}$ and $s^{++}$ pairing symmetries are given in Figs.~5(a)-5(c). 
For $s^{\pm}$-pairing, a spin-resonance
appears due to the inelastic scattering of the Bogoliubov quasiparticles whose energy and wave vector can approximately be determined
from $E=|\Delta^{\nu}_{{\bf k}_F}|+|\Delta^{\nu^{\prime}}_{{\bf k}_F+{\bf Q}}|$ given that sign$[\Delta^{\nu}_{\bf k}]\ne$sign[$\Delta^{\nu^{\prime}}_{{\bf k}+{\bf
Q}}]$ between band indices $\nu$ and $\nu^{\prime}$, where $\Delta^{\nu}_{\bf k}=\Delta^{\nu}_0g({\bf k})$ with $g({\bf k})=(\cos{k_x}+\cos{k_y})$. There are two reasons  for the  resonance shift to $E <2\Delta_0$:
(1) Due to large area of electron pocket in these systems \cite{Liu_arpes}, the effective gap value on the Fermi momenta is reduced,
i.e. $|g({\bf k}_F)|<1$. (2) The resonance energy shifts further to lower energy within RPA \cite{Dasresonance}. We take
$\Delta_0\approx 6$~meV \cite{Liu_arpes} to obtain a resonance at 7~meV, in accord with experimental value.
For the $s^{++}$ pairing, due to the lack of sign-reversal, the spin-excitations inside the SC gap are  completely eliminated. However, at
$E >2\Delta_0$, a hump-like feature in  intensity appears.
 The many-body RPA correction shifts
the hump to a higher energy (for the same value of $U=1.6$~meV, we obtain a weak feature around $E =1.3(2\Delta)$), as shown in
Fig.~5(b) and 5(c). With varying $U$ as well as the intrinsic broadening, we find that the result is robust and the $s^{\pm}$ resonance peak is ubiquitously sharper than that for the $s^{++}$-case. 
Therefore, our low-temperature neutron scattering results in Fig. 2 are only consistent with $s^{\pm}$-pairing symmetry
regardless of the actual values of the SC gaps \cite{Liu_arpes,thirupathaiah}.

\begin{figure}[t] \includegraphics[scale=0.45]{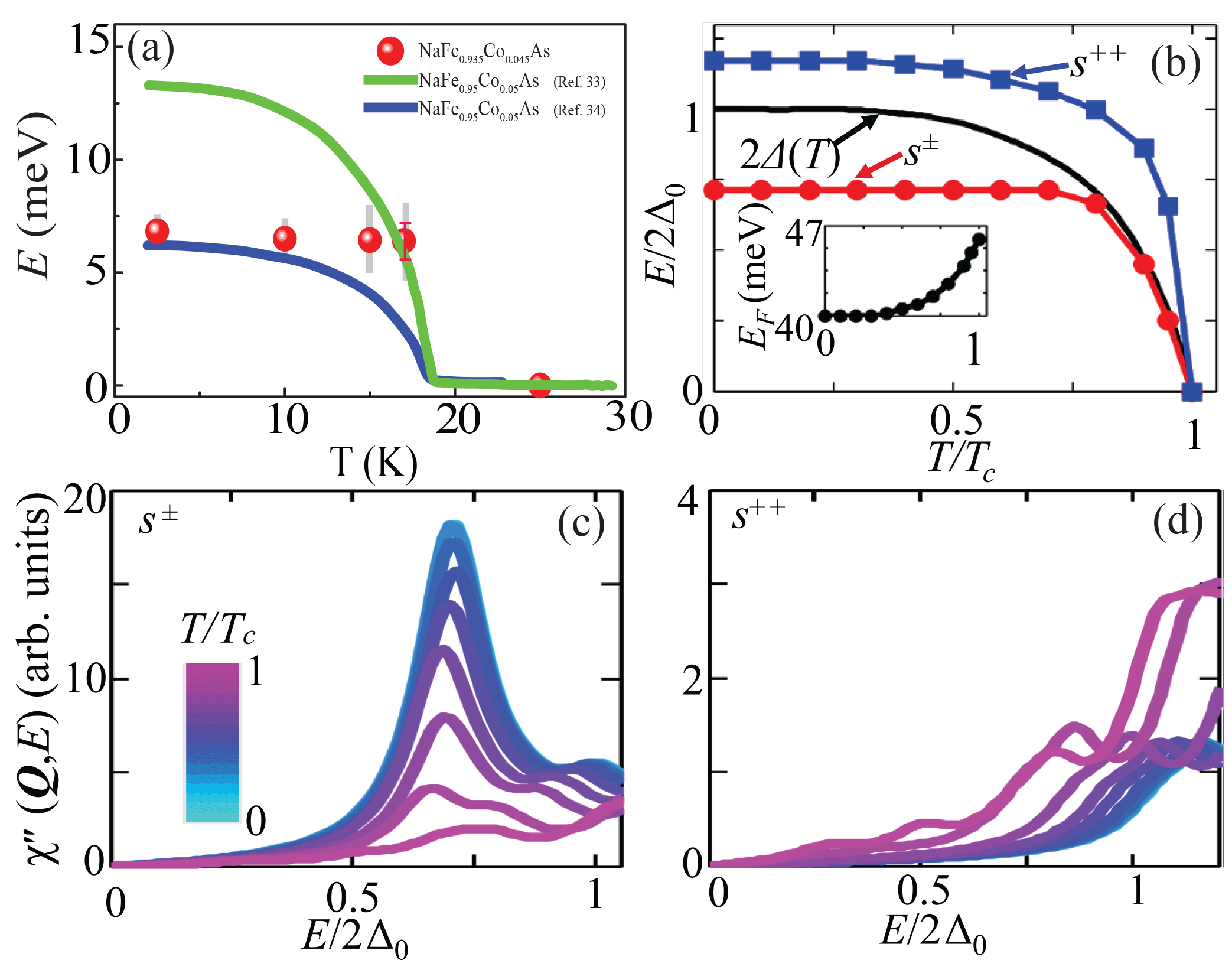} 
\caption{ (Color online)  Temperature evolution of spin-resonance for $s^\pm$ and $s^{++}$ pairing symmetries. (a) Temperature dependence of the peak position and energy width of the resonance, compared with temperature dependence of the SC gap energies as obtained from ARPES measurements (solid green \cite{Liu_arpes} and blue \cite{thirupathaiah} curves).  The vertical shaded bars are full-width-at-half-maximum of the resonance.  The error bars of the resonance peak energies are smaller than the symbol size for $T\leq 15$ K.  (b) Self-consistent values of the spin-resonance energy $E_{res}$ plotted as a function of temperature in the SC state for the two pairing symmetries under consideration, and compared with the SC gap amplitude.  All energy scales are normalized by the SC gap value at $T$=0, while the temperature is normalized by $T_c$ for presentation. The spin-resonance for $s^\pm$ pairing remains very much temperature independent except near $T_c$ due to self-consistently evaluated change in the Fermi level (shown in the inset) at each temperature, while that for the $s^{++}$ pairing follows the gap function at each temperature. (c) Spin-resonance peak and intensity as a function of energy at the commensurate wave vector at different temperatures (see the colorscale) for $s^\pm$ pairing symmetry. (d) Same as (c) but for the $s^{++}$ pairing symmetry.  Contrasting evolution of the resonance energy and intensity can be marked between $s^\pm$ and $s^{++}$ pairing, and our experimental data are consistent with the former pairing symmetry results.
}\label{fig6}
 \end{figure}

On the other hand, assuming that temperature dependence of the SC gaps in NaFe$_{0.935}$Co$_{0.045}$As follows the BCS form, as demonstrated in many pnictide superconductors \cite{borisenko,hirschfeld}, we observe that the resonance energy remains very much temperature-interdependent, see Fig.~6(a). To explain this behavior, we calculate the temperature evolution of the spin-susceptibility and the results are shown in Fig.~6(b-d). For each temperature, we evaluate the Fermi energy ($E_F$) constrained by the fixed number of electrons. As shown in inset to Fig. 6(b), $E_F$ increases as gap decreases with temperature. The corresponding change in the Fermi surface topology yields a change in the effective gap value for $s^\pm$ pairing due to the anisotropic gap structure factor $g({\bf k})$ defined before. Additionally, this leads to a reduction in $\chi^{\prime}$ in the $s^{\pm}$ pairing state, and for the temperature independent interaction $U$, the resonance condition shifts to a higher energy, as illustrated in the appendix. Due to the interplay between these two changes as a function of temperature, the resonance energy shifts to higher energy compared to its corresponding $2\Delta(T)$ value, but remains very much temperature-independent with respects to its $2\Delta(0)$ energy [Fig. 6(a)].  This analysis does not apply to the $s^{++}$ pairing symmetry since it is insensitive to the change in the underlying Fermi surface.  Therefore, the resonance energy in this pairing symmetry, which lies above $2\Delta(T)$, very much follows the gap function. 

The corresponding lineshape of the resonance is shown in Figs.~6(c) and 6(d) for $s^\pm$ and $s^{++}$, respectively. Contrasting temperature evolution of $\chi^{\prime\prime}$ at the commensurate wave vector can be marked between the two pairing symmetries. For $s^{\pm}$ pairing, the resonance intensity decreases according to the BCS prediction, but its position does not shift in energy until the temperature reaches the vicinity of $T_c$, where the resonance shifts to lower energy below its corresponding $2\Delta(T)$. On the contrary, for $s^{++}$ pairing the resonance intensity increases and gradually shifts to lower energy. This contrasting behavior can be understood from the physics of resonance described in the appendix.  
As shown in Fig. 8(b), the weak intensity hump above the $2\Delta$ in the $s^{++}$ arise from the discontinuous jump from the particle-hole channel to the SC region.  As gap decreases, the energy scale of the jump also decreases, and thus the 
resonance energy follows gap evolution, and remains almost insensitive to the modification in the underlying electronic structure.

In summary, we have discovered a sharp resonance in electron-overdoped NaFe$_{0.935}$Co$_{0.045}$As in the low-temperature SC state. our experimental data is consistent with the $s^\pm$ pairing state both in terms of 
 intensity and energy scales as a function of energy, and thus provide further evidence to the presence of $s^\pm$ pairing symmetry in pnictide superconductors.

\section{Acknowledgements} 
The single crystal growth efforts and neutron scattering work at UT/Rice are supported by the US DOE, BES,  through contract DE-FG02-05ER46202.  Work at IOP is supported by MOST (973 Project: 2012CB82400).  The work at JCNS and RWTH Aachen University was partially funded by the BMBF under contract No. 05K10PA3. Work at LANL was supported by the NNSA of the US DOE under contract DE-AC52-06NA25396.

\section{Appendix}

In this appendix, we first present elastic neutron diffraction data on the sample and then give the details of our calculations presented in the main text. The elastic neutron scattering scans across the AF Bragg peak positions are featureless and thus 
confirm this conclusion that the material does not have static AF order.

We evaluate the spin-resonance susceptibility in the superconducting (SC) state within the BCS-RPA formalism \cite{maier09,Dasresonance} which is given by
\begin{widetext}
\begin{eqnarray}
\chi_{BCS}^{rstu}({\bf q},\omega) &=& \frac{1}{N}\sum_{\bm k}M_{rstu}^{\nu,\nu^{\prime}}({\bm k},{\bm q}) \left\{\frac{1}{2}\left[1+\frac{\xi^{\nu}_{\bm k}\xi^{\nu^{\prime}}_{{\bm k}+{\bm q}} + \Delta^{\nu}_{\bm k}\Delta^{\nu^{\prime}}_{{\bm k}+{\bm q}} }{S^{\nu}_{\bm k}S^{\nu^{\prime}}_{{\bm k}+{\bm q}}}\right]\frac{f(S^{\nu}_{\bm k})-f(S^{\nu^{\prime}}_{{\bm k}+{\bm q}})}{\omega-S^{\nu}_{\bm k}+S^{\nu^{\prime}}_{{\bm k}+{\bm q}}+i\delta}\right.\nonumber\\
&& + \frac{1}{4}\left[1+\frac{\xi^{\nu}_{\bm k}}{S^{\nu}_{\bm k}}-\frac{\xi^{\nu^{\prime}}_{{\bm k}+{\bm q}}}{S^{\nu^{\prime}}_{{\bm k}+{\bm q}}}
-\frac{\xi^{\nu}_{\bm k}\xi^{\nu^{\prime}}_{{\bm k}+{\bm q}} + \Delta^{\nu}_{\bm k}\Delta^{\nu^{\prime}}_{{\bm k}+{\bm q}} }{S^{\nu}_{\bm k}S^{\nu^{\prime}}_{{\bm k}+{\bm q}}}\right]\frac{1-f(S^{\nu}_{\bm k})-f(S^{\nu^{\prime}}_{{\bm k}+{\bm q}})}{E+S^{\nu}_{\bm k}+S^{\nu^{\prime}}_{{\bm k}+{\bm q}}+i\delta}\nonumber\\
&&\left.+ \frac{1}{4}\left[1-\frac{\xi^{\nu}_{\bm k}}{S^{\nu}_{\bm k}}+\frac{\xi^{\nu^{\prime}}_{{\bm k}+{\bm q}}}{S^{\nu^{\prime}}_{{\bm k}+{\bm q}}}
-\frac{\xi^{\nu}_{\bm k}\xi^{\nu^{\prime}}_{{\bm k}+{\bm q}} + \Delta^{\nu}_{\bm k}\Delta^{\nu^{\prime}}_{{\bm k}+{\bm q}} }{S^{\nu}_{\bm k}S^{\nu^{\prime}}_{{\bm k}+{\bm q}}}\right]\frac{f(S^{\nu}_{\bm k})+f(S^{\nu^{\prime}}_{{\bm k}+{\bm q}})-1}{E-S^{\nu}_{\bm k}-S^{\nu^{\prime}}_{{\bm k}+{\bm q}}+i\delta} \right\}.
\label{seq:1}
\end{eqnarray}
\end{widetext}
Here $S^{\nu}_{\bm k}=[(\xi_{\bm k}^{\nu})^2+(\Delta_{\bm k}^{\nu})^2]^{1/2}$ is the SC quasiparticle state for the eigenstate $\xi_{\bm k}^{\nu}$ ($\nu$ is band index) and the SC gap $\Delta_{\bm k}^{\nu}$. For $s^{\pm}$-pairing symmetry we have $\Delta_{\bm k}^{\nu}=\Delta^{\nu}_0(\cos{k_x}+\cos{k_y})$ and for $s^{++}$-wave pairing symmetry we take $\Delta_{\bm k}^{\nu}=\Delta^{\nu}_0$ to be same for all bands. $M$ is the matrix element consists of the eigenstates of the initial and final scattered quasiparticle state as given by $M_{rstu}^{\nu,\nu^{\prime}}({\bm k},{\bm q})=\phi_{r}^{\nu^{\prime}\dag}({\bm k}+{\bm q})\phi_{s}^{\nu}({\bm k})\phi_{t}^{\nu\dag}({\bm k})\phi_{u}^{\nu^{\prime}}({\bm k}+{\bm q})$, where $\phi_{r}^{\nu}({\bm k})$ is the eigenfunction of the $\nu^{th}$-band derived from $r^{th}$-orbital.

\begin{figure}[t] \includegraphics[scale=.35]{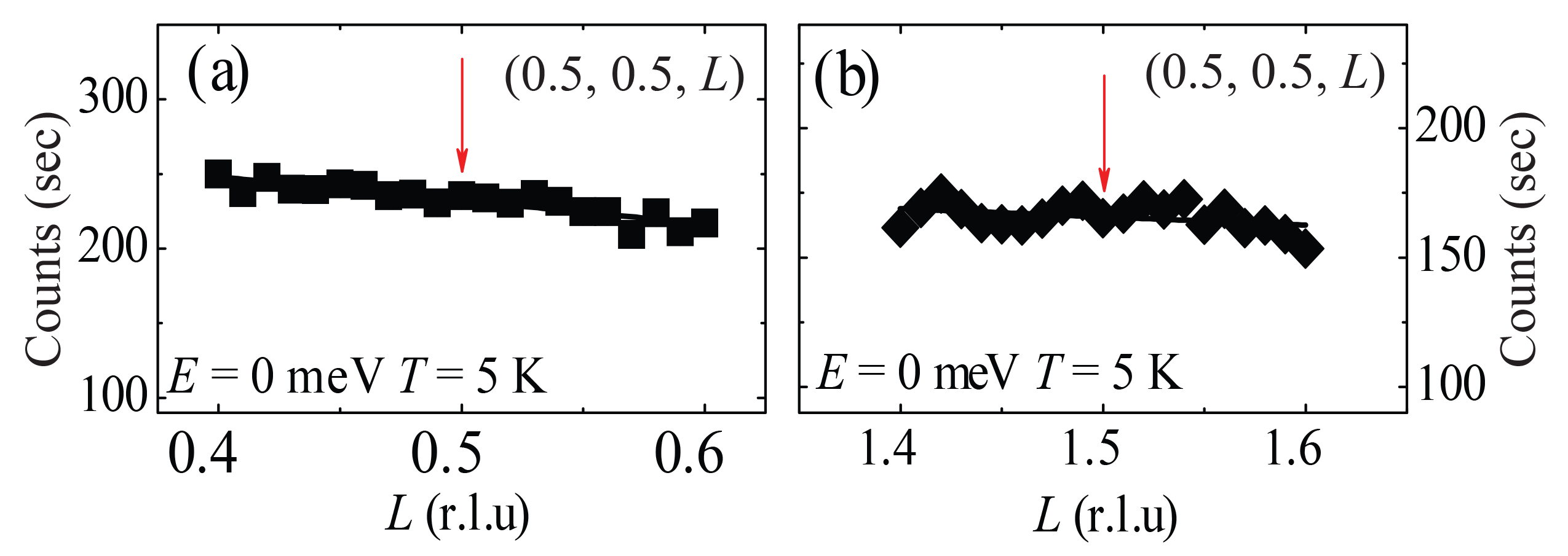} 
\caption{{\bf Elastic neutron scattering through the AF Bragg positions.} (a) and (b) Elastic neutron scattering from PUMA along the $[0.5,0.5,L]$ directions at 5 K.  The solid lines are guided to the eyes and the arrows indicate the positions of
 AFM static ordering.}
\end{figure}

In Eq.~\ref{seq:1} in the appendix, the first term is called particle-hole scattering term which vanishes in the SC state due to particle-hole symmetry regardless of any pairing symmetry. The second and third term are called particle-particle and hole-hole scatterings, respectively which become active in the SC state. Focusing on the third term (the same analysis applies to the second term), we find that this term contributes a non-zero value only when sign$[\Delta^{\nu}_{\bm k}]\ne$sign$[\Delta^{\nu^{\prime}}_{{\bm k}+{\bm q}}]$ (since $\xi^{\nu}_{\bm k} =0$ on the Fermi surface). A pole thus obtained in the imaginary part of $\chi_{BCS}$ at
\begin{eqnarray}
E= |\Delta^{\nu}_{\bm k}|+|\Delta^{\nu^{\prime}}_{{\bm k}+{\bm q}}|.
\end{eqnarray}
(Of course, the many body effect and the matrix-element effect can shift the energy scale as discussed below). For $s^{++}$ pairing, the lack of sign-reversal prevents any resonance to occur inside the gap. Here, for $\omega>2\Delta$ Eq.~\ref{seq:1} transform into the particle-hole (p-h) channel governed by the free-fermion Lindhard function as
\begin{eqnarray}
\chi_{0}^{rstu}({\bm q},E) = \frac{1}{N}\sum_{\bm k}M_{rstu}^{\nu,\nu^{\prime}}({\bm k},{\bm q}) \frac{f(\xi^{\nu}_{\bm k})-f(\xi^{\nu^{\prime}}_{{\bm k}+{\bm q}})}{E-\xi^{\nu}_{\bm k}+\xi^{\nu^{\prime}}_{{\bm k}+{\bm q}}+i\delta}.
\end{eqnarray}
Unlike the BCS susceptibility, the p-h continuum is considerably broad and is obtained from the dynamical nesting between different bands as $E_{q}=\xi_{\bm k}^{\nu}-\xi_{{\bm k}+{\bm q}}^{\nu^{\prime}}$.  For $s^{\pm}-$pairing, $\chi_{BCS}$ gives a very sharp resonance feature inside the SC region, which virtually washes out the information of the p-h continuum above this energy scale. On the other hand, for $s^{++}$-pairing, $\chi_{BCS}=0$ and the sharp crossover to $\chi_0$ at $E=2\Delta$ acts as an abrupt change in the real part of $\chi_0$. In both cases, the Fermi surface nesting between electron-pockets sitting at M point and the hole-pockets at $\Gamma$-point is most dominant, giving features at the commensurate wave vector ${\bm Q}=(\pi,\pi)$. The characteristic differences in the bare susceptibilities for the two pairing channels are given in the supplementary Fig.~2.

\begin{figure}[t]
\includegraphics[scale=.35]{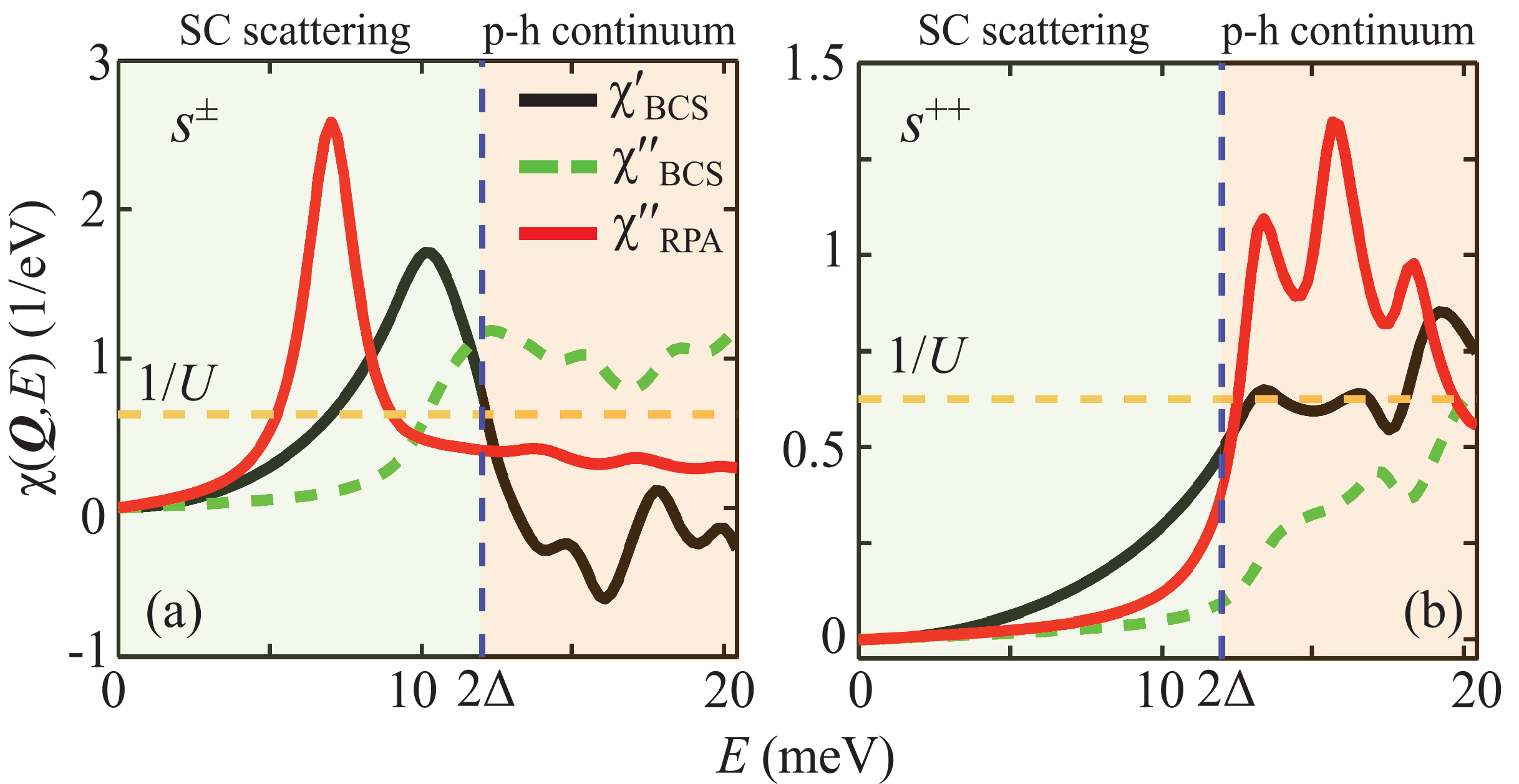}
\caption{ (a) Various components of the spin-susceptibility are plotted together for $s^{\pm}$-pairing. The RPA value of the susceptibility (red line) is very sharp (of the order of 100 eV$^{-1}$), and thus we divide it by 30 to fit it into the same figure or visualization. For the same reason, the real part of the bare susceptibility $\chi_{BCS}^{\prime}$ is subtracted from its zero energy value and multiplied by 4.9 and the imaginary part of the BCS susceptibility is multiplied by 2. (b) Same as (a) but for $s^{++}$-pairing. Here $\chi_{RPA}^{\prime\prime}$ is divided by 20 and $\chi_{BCS}^{\prime\prime}$ is multiplied by 2. The horizontal dashed line gives $1/U$, which illustrated the occurrence of the resonance mode when it cuts through $\chi_{BCS}^{\prime}$. Finally, the vertical dashed line marks the $2\Delta$ line which separates the SC scattering region to the p-h continuum.}
\end{figure}

We include the many-body correction to the spin-excitation spectrum via standard RPA formalism as $\tilde{\chi}_{RPA}=\tilde{\chi}_{BCS}/(\tilde{1}-\tilde{U}\tilde{\chi}_{BCS})$, where the symbol tilde over a quantity represents that it is a matrix in the orbital basis. The interaction vertex consists of intra-orbital Coulomb interaction $U$, inter-orbital term $V=2U^{\prime}-J$, where $U^{\prime}=U-5/4J$ and $J=U/8$ is the Hund's coupling and $J^{\prime}=J/2$ is the pair hopping term. For a single band system, a spin-resonance occurs when the denominator of RPS susceptibility vanishes at $\chi_{BCS}^{\prime}=1/U$, and the peak is broadened by $\chi_{BCS}^{\prime\prime}$. The situation is equivalent, yet complicated in the multi-orbital case. We obtain a sharp resonance peak for the $s^{\pm}$-pairing case below $2\Delta$ where the broadening associated with $\chi_{BCS}^{\prime\prime}$ is much reduced, as shown in supplementary Fig.~1(a). On the other hand, for $s^{++}$ the resonance in the p-h continuum is broadened by large value of $\chi_{0}^{\prime\prime}$ [see green dashed line in supplementary Fig.~1(b)].


\end{document}